\documentclass[prl,aps,showpacs,twocolumn,floats,superscriptaddress]{revtex4}
\usepackage{amsmath} 
\usepackage{amssymb} 
\usepackage{graphicx} 
\usepackage{epsfig} 
\newcommand{\beq}{\begin{equation}} 
\newcommand{\eeq}{\end{equation}} 
\newcommand{\bea}{\begin{eqnarray}} 
\newcommand{\eea}{\end{eqnarray}}

\begin{document}
\title{Toward a microscopic description of dimer adsorbates on  metallic surfaces}
\author{Jaime Merino}
\affiliation{Departamento de F\'\i sica Te\'orica de la Materia Condensada, Universidad 
Aut\'onoma de Madrid, Madris 28049, Spain}
\author{L\'aszl\'o Borda}
\affiliation{Physikalisches Institut and Bethe Center for Theoretical Physics, Universit\"at Bonn, Nussallee 12 D-53115 Bonn, Germany}
\author{Pascal Simon}
\affiliation{Department of Physics and Astronomy, University of Basel,
CH-4056 Basel, Switzerland}
\affiliation{Laboratoire de Physique et Mod{\'e}lisation des Milieux
  Condens{\'e}s, CNRS and Universit{\'e} Joseph Fourier, BP 166, F-38042
  Grenoble, France}
\date{\today}
\pacs{72.15.Qm, 68.37.Ef, 72.10.Fk,75.30.Hx}

\begin{abstract} 
Despite the experimental successes of Scanning Tunneling Microscopy (STM) and the interest in more complex
magnetic nanostructures, our present understanding and theoretical description
of STM spectra of magnetic adatoms is mainly phenomenological and most often ignores quantum many-body effects.
Here, we propose a theory which includes a microscopic description 
of the wave functions of the substrate and magnetic adatoms together with
the quantum many-body effects. 
To test our theory, we have computed the  
STM spectra of magnetic Cobalt monomers and dimers adsorbed on a metallic Copper surface 
and successfully compared our results 
to recent available experimental data.
\end{abstract} 
\maketitle 
 
{\em Introduction}
In the past decade, STM experiments of magnetic adsorbates 
on metal surfaces have florished. They have identified sharp resonances at the Fermi surface \cite{madhavan98,berndt98,corral1,knorr,limot07},
characteristic of the Kondo effect, a paradigmatic quantum many-body state which originates from the screening
of the impurity spin by the conduction electrons. Systematic experimental studies of more complex structures
with two or more magnetic atoms
have been recently initiated 
and strong magnetic exchange interactions between the impurities 
have been identified \cite{chen99,jamneala} and estimated \cite{heinrich,wahl07}. 
Despite all these important experimental successes  and the interest in more complex
magnetic nanostructures, our present understanding
is mainly phenomenological. 
Since the exponentially small Kondo energy scale cannot be captured
by present-day  calculations based on the Density Functional Theory (DFT),
a full ab initio theory of Cobalt impurities on metallic surfaces is precluded.
A different strategy is therefore required. Here, we proceed in two separate steps by
 combining two complementary approaches:
the single particle physics which is mainly responsible for the shape of the STM spectra
is described by solving the Schroedinger equation in an effective  potential which 
parameterizes the metallic surface
and the adatoms
while the quantum many-body physics is treated separately 
via the powerful numerical renormalization group (NRG) method.
Such strategy captures low energy physics and should therefore 
be suitable to describe STM spectra of magnetic adatoms around 
zero bias. This
constitutes a first step toward a microscopic  description of competing quantum 
many-body states at metallic surfaces. 

The Kondo effect  
occurs when the spin of  magnetic adsorbate couples to the surrounding electrons. Below 
a characteristic energy scale $T_K$, named the Kondo temperature, the conduction electrons  
form a strongly-correlated singlet with  
the spin of the adsorbate which results in its screening. In STM experiments, this manifests as 
a  zero-bias resonance of width $T_K$ in the differential conductance\cite{madhavan98,berndt98,corral1,knorr,limot07}. The theoretical description of the STM spectrum of a single magnetic atom adsorbed on a metallic surface   
is based on a Fano line shape \cite{Fano} associated with a {\it non-interacting} resonant level.
The Fano function is most often  empirically 
assumed  while its parameters are adjusted phenomenologically \cite{schiller,ujsaghy,plihal,madhavan01} or
computed from a microscopic description of metal-adsorbate-tip interactions \cite{merino,neto06}.

When two magnetic adsorbates or more are brought into proximity, magnetic interactions 
between the impurities --direct or indirect-- 
start competing with the independent screening of each impurity. From this competition different magnetic 
ground states can be reached \cite{jones87,jones88,silva96} that can strongly affect the differential conductance \cite{simon05} and may  
even lead to non-Fermi liquid behavior in small clusters \cite{lazarovits}.  
In Ref. \cite{wahl07}, Wahl {\it et al.} perform a systematic experimental analysis of Cobalt dimers adsorbed on a Copper surface by increasing the interatomic distance between the adatoms and therefore decreasing their magnetic exchange interactions.
The measured changes of the STM spectra with the adatom interatomic distance thus 
permitted an indirect estimate of this   
coupling via STM. Below, we propose a  theoretical description of low energy STM spectra  
 of magnetic monomers and dimers adsorbed on a metallic surface by combining microscopic
evaluation of the one-electron parameters with NRG 
calculations without making any phenomenological assumptions on the spectral shapes.
We focus on Cobalt atoms adsorbed on Copper (100) to compare our results with recent experimental data by Wahl {\it et al.} \cite{wahl07} but our approach can also be extended to other magnetic adatoms and metallic surfaces.

{\em Model} 
We consider a cluster of $N_c$  magnetic atoms 
adsorbed on a metallic surface. The complete system involves the STM tip, the substrate 
and can be described by the following Hamiltonian  
\begin{equation} 
H=H_{subs}+H_{tip-subs}+H_{tip}, 
\label{model}
\end{equation} 
where $H_{subs}$ describes the substrate plus the adsorbed 
3d transition metal atom and $H_{tip-subs}$ describes the 
interaction of the tip with the substrate.  $H_{tip}$ describes 
the tip which is assumed to have an unstructured density of states. 
The substrate with the adsorbed 3d atom may be 
modeled by a generalized Anderson model. 
The $d_{3z^2-r^2}$ orbital is the most strongly coupled
to the metal at typical adsorption distances
of Co on Cu(100) due to the lobe pointing to 
the surface which overlaps more strongly with 
the metal wavefunctions. The d-orbitals parallel to 
the surface x-y $(d_{xy}, d_{x^2-y^2})$ plane lead to cancellations between the
negative and positive lobes of the adsorbate orbital. This has been  checked 
by evaluating the matrix elements along  the lines shown in Ref. [15]. 
We therefore assume in the sequel that only the  $d_{3z^2-r^2}$ orbital of the adatoms hybridizes with the 
substrate \cite{Weissman}.

The metallic states, denoted by $|{\bf k}>$, couple to the 
$d_{3z^2-r^2}$ orbitals 
denoted by $|d_i>$ with $i=1,.., N_c$  labeling each adatom of the cluster. 
The Anderson model usually assumes that the continuum of 
metal states are orthogonal to the localized orbitals of the 
adsorbate. However, the basis set formed by the unperturbed metal, 
adsorbates and tip states, $\{ |{\bf k}>, |d_i>, |t> \}$, is 
generically non-orthogonal  and over-complete. 
One way to correct this 
is to redefine the metallic states, ${\bf k}$, as:\\
$
|{\bf \tilde{k} }> = |{\bf k}> - \sum_i <d_i|{\bf k}>|d_i>-<t|{\bf 
k}>|t>. \label{ortk} 
$
Considering that tip and adsorbate wavefunctions are orthogonal: 
$<t|d_i>=0$, as $|d_i>$ is very localized, then the new metallic 
states satisfy 
$ 
<\tilde{{\bf k}}|\phi>=0, 
$ 
where $|\phi>$ can be either $|t>$ or $|d_i>$. 
 
Our starting point is thus an Anderson model defined in this new 
orthogonal basis: $\{ |{\bf \tilde{k}}>, |d_i>, |t> \}$, from 
which associated one-electron parameters
are obtained.  An analogous procedure was previously used in the 
context of chemisorption of atoms and molecules on metal surfaces 
by Grimley \cite{Grimley}. In this basis, the substrate Hamiltonian reads 
\begin{eqnarray} 
&&H_{subs}=\sum_{{\bf k}\sigma } \epsilon_{\bf \tilde{k} } 
c^{\dagger}_{{\bf \tilde{k}}\sigma} c_{{\bf \tilde{k}}\sigma} + 
\epsilon_d \sum_{j\sigma} d^{\dagger}_{j\sigma} d_{j\sigma} 
 \\ 
&&+ \sum_{j=1}^{N_c}\sum_{ {\bf k},\sigma} V_{{\bf \tilde{k}}j} 
(d^{\dagger}_{j\sigma} c_{{\bf \tilde{k}}\sigma} + H. c.) 
+U \sum_{j,\sigma < \sigma'} d^{\dagger}_{j\sigma} d_{j\sigma} 
d^{\dagger}_{j\sigma'} d_{j\sigma'}.\nonumber  \label{subs} 
\end{eqnarray} 
Here $\epsilon_d$ is the energy level of an electron residing in 
the $d_{3z^2-r^2}$ orbital of the adsorbate, $c^{\dagger}_{{\bf \tilde{k}}m 
\sigma}$ creates an electron with spin $\sigma$, momentum ${\bf 
\tilde{k}}$ in the metal. $d^{\dagger}_{j\sigma}$ creates an 
electron with spin $\sigma$ in adsorbate-$j$. $\epsilon_{\bf 
\tilde{k}}$ and $V_{{\bf \tilde{k}}j}$ are the metallic energies 
and the hybridization matrix elements between the substrate and 
the adsorbate-$j$, respectively (notice that  
the momentum dependence of the hybridization matrix 
elements is explicitly taken into account). $U$ is the Coulomb repulsion of 
two electrons in the 3d orbital of the transition metal atom. 
Finally, the tip-substrate interaction contribution to the 
Hamiltonian reads 
\begin{equation} 
H_{tip-subs}=\sum_{{\bf k},\sigma} M_{ {\bf \tilde{k}}} 
( c^{\dagger}_{{\bf \tilde{k}}\sigma } t_{\sigma} + H. c.), 
\label{tip} 
\end{equation} 
through the matrix elements, $M_{{\bf \tilde{k}}}$. Here, 
$t_{\sigma}$ destroys an electron with spin $\sigma$ 
in the tip.  We have neglected the 
direct coupling of the tip with the substrate d bands and with the 
3d orbital of the adsorbates due to the localized nature of these 
d orbitals.

{\em Hybridization matrix elements} 
In the following we briefly describe how hybridization matrix elements, 
$M_{{\bf \tilde {k}}}$ and $V_{{\bf \tilde{k}}}$, are computed. 
For simplicity we first focus on how $V_{\bf \tilde{k}}$ is 
computed, as $M_{{\bf \tilde{k}}}$ is computed in a similar way. 
We first re-express the Hamiltonian in Eq. (\ref{subs}) 
in first quantized form 
$ H_{subs}=T+\sum_{j=1}^{N_c}V_j+V_M,$  
where $T$ is the kinetic energy of the system, $V_j$ is the 
potential created by adsorbate-$j$ and $V_M$ describes the surface 
potential. 
The matrix elements between the 
orthogonalized metallic states, ${\bf \tilde{k}}$, and the adsorbate 
thus read: 
\begin{equation} 
V_{{\bf \tilde{k}}j}= V_{{\bf k}j}-S_{{\bf k}j}<d_j|V_M|d_j>, 
\label{matrixel} 
\end{equation} 
where, again, we have assumed: $<t|d_j>=0$ and 
$<d_i|d_j>=\delta_{ij}$.  
The first term in Eq.~(\ref{matrixel}) is 
the hybridization matrix element with the 
unperturbed wavefunctions ${\bf k}$: 
$ 
V_{{\bf k}j} = <{\bf k}|V_M|d_j>, 
$ 
and the second contains the overlap matrix element 
$ 
S_{{\bf k}j} = <{\bf k}|d_j>$. 
The above orthogonalization procedure 
automatically selects the metal potential $V_M$ in the 
hybridization matrix elements favoring the region close to or inside the metal 
in the integrations. This differs from hybridization matrix elements 
computed with the original wavefunctions as in that case integrations 
over the whole space are involved.  
The metal effective potential for the Cu(100) surface, $V_M$, is parameterized based on 
the Jones-Jennings-Jepsen (JJJ) potential \cite{Jones} (see Ref. \cite{merino} for 
details).  
The sharpness of the surface barrier potential $\lambda=2.2 $\AA$^{-1}$, and the image  
plane position of $Z_{im}=1.15$ \AA.  The adsorbtion distance of Co on  
Cu(100) is estimated to be $Z_d=1.5$ \AA.  
Finally we note that orthogonalization effects enter the 
model through matrix elements only. Orthogonalization effects on 
the substrate band energies can be shown to be of higher order in the 
overlap. Hence, we assume $\epsilon_{\bf \tilde{k}}=\epsilon_{\bf k}$ in the rest of the paper.

{\em STM conductance results}
Following Ref.~\cite{madhavan01} and for the sake of clarity 
we derive the basic equations needed for the computation of the 
conductance through the STM. If we neglect any modification of the 
substrate due to the presence of the tip (this is reasonable 
considering the fact that the tip is typically at about $5-10$ \AA~above the metal surface), 
then the differential conductance measured by the STM 
reads: \cite{schiller,plihal,madhavan01} 

\begin{equation} 
\frac{dI}{dV }(\omega) = 
\frac{4 e^2}{\hbar} \rho_{tip} \int d\epsilon 
\frac{ \partial f(\epsilon - \omega)}{\partial \epsilon } 
(\Gamma(\epsilon) + 
\delta \Gamma(\epsilon)), \label{conduct} 
\end{equation} 
where 
$ 
\Gamma(\omega) = \pi \sum_{\bf k}|M_{{\bf \tilde{k}}}|^2 
\delta(\omega-\epsilon_{\bf k}) 
$ 
is the conductance associated with the clean substrate (without the 
adsorbed 3d transition metal atom) and $\rho_{tip}$ is the density of states of 
the tip and $f(\epsilon)$ the Fermi-Dirac function.
Modifications of the tip-surface coupling induced by the presence of the 
adsorbate are given by 
\begin{equation} 
\delta \Gamma(\omega)= {\rm Im} \sum_{i,j}\sum_{{\bf k, k'}} \frac{M_{ {\bf 
\tilde{k} } } V_{ {\bf \tilde{k} }i}}{ \omega - \epsilon_{\bf 
k} - i \eta} G_{ij}(\omega) \frac{ V^*_{j {\bf \tilde{k'}}} M^*_{ {\bf 
\tilde{k'}} } }{ \omega - \epsilon_{\bf k'} -i \eta}. 
\label{gamma} 
\end{equation} 
The Greens function, $G_{ij}(\omega)$, describes the electronic 
properties of the cluster of $3d$ adsorbates immersed in the 
metallic continuum including the quantum many-body effects such as the 
Kondo effect but also the interactions between the Co adatoms.
  $\eta$ is an analytical continuation parameter. 
For convenience Eq.~(\ref{gamma}) is rewritten in the following way 
$
\delta \Gamma(\omega) = {\rm Im}  \sum_{i,j}\{ (A_i(\omega) + i B_i(\omega)) 
G_{ij}(\omega) (A_j^*(\omega)+i B_j^*(\omega)) \} \label{gammaAB} 
$
where $B_i(\omega)=\pi\sum_{{\bf k}}M_{{\bf \tilde k}} V_{{\bf \tilde k}} \delta(\omega-\epsilon_{\bf 
k})$  and $A_i(\omega)$ 
is the Kramers-Kronig transformation of $B_i(\omega)$.  
For the systems of interest here, $A_i(\omega)$ and $B_i(\omega)$ are 
real. Function $B_i(\omega)$ embodies the information concerning 
the tip-substrate-adsorbate system as it depends on the 
tip-adsorbate separation, on the position of the adsorbate 
with respect to the last surface plane of ions, and on the metal potential 
described by $V_M$ through the matrix elements, $V_{{\bf \tilde{k}}i}$ and $M_{\bf \tilde{k}}$.  
In the single impurity case, $dI/dV(\omega)$ reduces to a Fano form\cite{Fano} for $\omega<T_K$ and 
the line shape is determined by the Fano parameter\cite{merino}: $q=A(0)/B(0)$, 
evaluated at the Fermi energy taken as zero. Thus, the shape is fixed by the one-electron
parameters entering the full model (\ref{model}).  
 
We apply this formalism to compute the conductance of  two Co atoms deposited 
on Cu(100) surfaces. We take in the following the origin of distances 
at one of the Co adatoms in the dimer ${\bf R_1}=0$ and the distance  
between the atoms is denoted by $d$. The magnetic exchange interaction
depends on the interatomic distance $d$ and is denoted as $I(d)$. This function
has been calculated using {\it ab initio} calculations by Stepanyuk et 
al.\cite{stepanyuk}  
and also estimated experimentally \cite{wahl07}.  Due to the small spatial range of the d orbitals compared to the typical separation between the adatoms considered here, this magnetic
exchange interaction is mainly of RKKY type as it has been shown in \cite{stepanyuk}.
For our purpose, we merely regard $I(d)$ 
as an external parameter which may eventually extracted from ab initio calculations. 
 
The challenging task is to compute the two by two matrix $G_{ij}$ in Eq. (\ref{gamma}) which encodes
interaction effects. In order to perform this task, we use the powerful NRG approach.
However, a full NRG treatment of
two five-fold degenerate impurities is a way beyond today's computational possibilities.
We choose instead a simpler route and 
first map the Anderson Hamiltonian in Eq. (\ref{subs}) 
to a Kondo-like Hamiltonian. Such procedure is standard and justified by the fact 
that the electron in the $d_{3z^2-r^2}$ orbital of the Cobalt adatom is strongly bound.
The resulting Kondo model is described by the two-impurity Kondo Hamiltonian \cite{jones87,jones88}.  We neglect potential scattering terms since they do not renormalize
and barely  contribute to the low energy physics. 
The low energy physics of the 
two-impurity Kondo model is then controlled by only two
energy scales: the single-impurity Kondo temperature $T_K$
and the magnetic exchange interaction $I$ \cite{jones87,jones88}.
We use NRG calculations to  obtain the impurity spectral properties. 

Wilson's NRG technique \cite{Wilson} is a non-perturbative and 
numerically exact method suitable especially for quantum impurity problems (for a recent review see \cite{bulla}). 
The cornerstone of the method is the logarithmic discretization  of the 
conduction electron band and mapping the system onto a 
semi-infinite chain with the impurity at the end.  
NRG has been successfully applied to the problem of two magnetic  
impurities\cite{jones88,silva96} 
but in those earlier works only thermodynamic 
quantities were computed.  
The main source of complication in a two impurity calculation 
as compared to a single impurity case is that the former is an 
effectively two band calculation: From NRG point of view  
it is 
a challenging task because the impurities now couple to two semi-infinite 
chains. Consequently, the Hilbert space grows by a factor 
16 in each NRG step. This is still manageable  
with today's computer resources. 
 Concerning the details of the present NRG calculation, the
conduction and is discretized logarithmically in intervals $[\Lambda^{n+1}D,\Lambda^n D]$
and $[-\Lambda^{n} D,-\Lambda^{n+1}D]$ where $D$ is the bandwidth and $\Lambda$ 
the discretization parameter.
We took
 $\Lambda=2$,  the number of iterations  
was $N=60$ 
and we kept $M=3072$ states 
per iteration exploiting the charge and spin $z$-component $U(1)$ 
symmetries. The calculations were performed at $T=0$ and $I(d)$ was taken as an input parameter we took from  
{\em ab initio} calculations \cite{stepanyuk}. 
In the NRG calculations, we have neglected the details of  
the substrate band structure and the $k$-dependence of the 
matrix elements $V_{kj}$ since the low energy properties of the two impurity Kondo model 
mainly depend  
on the single impurity Kondo temperature and the RKKY interaction. Therefore, we expect  
our whole approach combining single particle description of the adsorbate together 
 with NRG calculations 
to be qualitatively correct at low energies {\it i.e.} around zero bias and within an interval of order a few $T_K$. It would be interesting but quite computationally demanding to apply 
NRG techniques directly on the two-impurity Anderson Hamiltonian neglecting
orbital degeneracy with the matrix elements $V_{kj}$ computed from the previous 
single-particle calculations. We leave this for future work.

We would like to note here that in principle NRG is capable of
reproducing RKKY interaction, see Ref.\cite{silva96}. Here we choose a
different strategy, by mapping the problem onto two independent Wilson chains,
but in contrast to Ref.\cite{jones88}, we take the finite overlap between
the electron fields centered around the positions of the two impurities into
account. In this way NRG does not generate additional RKKY interaction thus
the danger of double-counting is avoided. The reason why we follow this
strategy is that the RKKY interaction, generated in the NRG scheme of 
Ref.\cite{silva96} for a Kondo-like model with constant, featureless
conduction electron DoS, has a very little to do with the realistic $I(d)$
obtained by {\em ab initio} method. 

\begin{figure} 
\begin{center} 
\epsfig{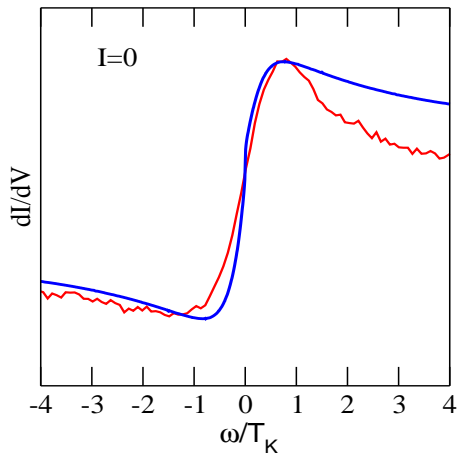} 
\end{center} 
\caption{Kondo resonance for a single Co adatom on a Cu(100) surface. 
STM measurements at T=6 K (red line) are compared to our theoretical 
results (black line). 
}\label{fig1} 
\end{figure} 
When the two Cobalt adatoms are far apart, the RKKY interaction is negligible and the adatoms 
can be regarded as isolated. We have plotted in Fig. \ref{fig1} 
the differential conductance
-function of the dimensionless parameter $\omega/T_K$- 
calculated using the aforementioned procedure for a single Co adatom and compared it to experimental 
raw data. The agreement is extremely good at low energy 
and qualitatively correct at higher energy. 

\begin{figure}
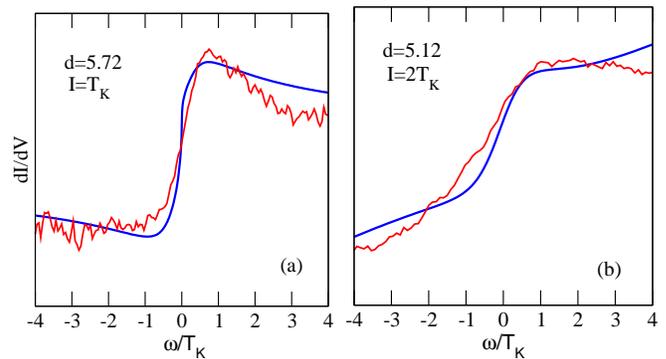
 
\begin{center} 
\epsfig{file=ldosI1TK_EXPT.eps,width=4.30cm,angle=0,clip=} 
\epsfig{file=ldosI2TK6_EXPT.eps,width=4.20cm,angle=0,clip=} 
\end{center} 
\caption{Kondo resonance for Co antiferromagnetically coupled Co dimers on a Cu(100) surface. 
The distance, $d$, (in \AA) between the Co atoms is reduced from (a) to (b) with the 
corresponding coupling $I(d)$ \cite{stepanyuk} increasing as shown. 
In (a) and (b) we compare theoretical calculations for an antiferromagnetic 
interaction, $I>0$, and available experimental data \cite{wahl07} at 6 K. 
In (b) a linear background present in the experimental data has been 
appropriately added to the theoretical spectra for comparison. 
} 
\label{fig2} 
\end{figure}

\begin{figure}
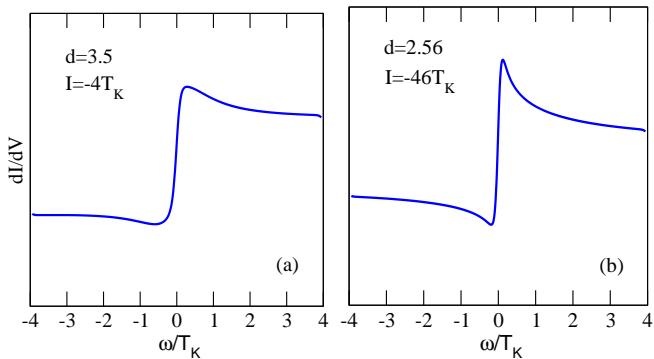

\begin{center}
\epsfig{file=ldosI-4TK_NEW.eps,width=4.30cm,angle=0,clip=}
\epsfig{file=ldosI-46TK_NEW.eps,width=4.20cm,angle=0,clip=}
\end{center}
\caption{Kondo resonance for ferromagnetically coupled dimers on a Cu(100) surface.
Experimental data do not show a Kondo effect at 6 K for d=2.56 \AA,  which is in agreement 
with the strong reduction in the Kondo temperature predicted here.
}
\label{fig3}
\end{figure}

Let us now apply the same method to obtain STM spectra for Cobalt dimers separated by a distance $d$ 
such that the RKKY interaction become of the order of $T_K$ or larger. The resulting differential conductance
on top of one adatom are summarized 
in Figs. \ref{fig2} and \ref{fig3} for various distances separating the Cobalt adatoms. 
According to {\it ab initio} results \cite{stepanyuk},  
we can associate to the distance d=5.72 \AA~  and d=5.12 \AA~  
the  antiferromagnetic RKKY exchange interactions $I\sim T_K\approx 7.6~meV$ and  $I\sim 2T_K\approx 15.6~meV$ respectively. We have compared in Fig. \ref{fig2} our results for the  
conductance on top of one adatom for these two distances to the experimental 
raw data obtained at $T=6K$  
with a very good agreement especially at low energies. The agreement is quite remarkable
 since  
the theory contains a single fit parameter: the magnetic exchange interaction $I$ 
which is extracted from {\it ab initio} calculations \cite{stepanyuk}). 
The correct shape of the STM signal is reproduced without using any Fano line shape fits 
which would be actually unjustified for Cobalt dimers. 
We then apply our method in order to predict  the differential conductance 
for the adatom separation corresponding to d=3.5 \AA~
and d=2.56 \AA~ as shown in Fig. \ref{fig3}. 
The magnetic exchange interaction is predicted to be ferromagnetic \cite{stepanyuk}
and corresponds to 
$I\approx -4T_K$) and $I\approx -46T_K$ respectively. In the latter case, the Kondo temperature is
predicted to be very small, $\approx 0.23$ meV and although is not accessible with present
STM experiments it is the predicted line shapes resulting from our combined approach at
very low temperatures. 

In summary, we have presented theoretical predictions for the STM spectra of magnetic monomers and dimers
adsorbed on metals. Our approach relies on a microscopic description of 
the adsorbate-metal-tip interactions combined with NRG calculations. 
Such methodology may constitute the first building block for 
understanding STM spectra of more complex magnetic structures on metal
surfaces. 

{\bf Acknowledgment} We acknowledge P. Wahl and K. Kern for discussions  and 
for sending us their raw data files for the Cobalt dimers adsorbed on Cu(100).
J.M. thanks O. Gunnarsson for useful comments on the manuscript. 
L.B. acknowledges the financial support of the Alexander von Humboldt
Foundation, J\'anos Bolyai Foundation and Hungarian Grants OTKA
through projects T048782 and K73361.

\end{document}